\documentclass[a4paper, 10pt]{article}

\newcommand\draft[1]{}
\newcommand\release[1]{#1}

\newcommand\bigone[1]{}
\newcommand\smallone[1]{#1}

\newcommand\relativepath{}

\newcommand{\cD}{{\cal D}}
\newcommand{\cH}{{\cal H}}

\draft{\usepackage{easy-todo}}

\draft{\usepackage[notref,notcite]{showkeys}}

\usepackage{graphicx}

\usepackage{amsfonts}
\usepackage{amssymb}
\usepackage{amsmath}
\usepackage{amsthm}
\usepackage{latexsym}

\bigone{\newtheorem{thm}{Theorem}[chapter]}
\smallone{\newtheorem{thm}{Theorem}}
\newtheorem{lem}[thm]{Lemma}

\theoremstyle{definition}

\newcommand{\refthm}[1]{Theorem~\ref{thm:#1}}
\newcommand{\reflem}[1]{Lemma~\ref{lem:#1}}

\newcommand{\refsec}[1]{Section~\ref{sec:#1}}

\newcommand{\refeqn}[1]{(\ref{eqn:#1})}

\newcommand{\refalg}[1]{Algorithm~\ref{alg:#1}}

\def \rf(#1:#2){\csname ref#1\endcsname{#2}}

\newcommand{\pfstart}{\begin{proof}} 
\newcommand{\pfsketch}{\begin{proof}[Proof sketch]}
\newcommand{\pfend}{\end{proof}} 
\newcommand{\itemstart}{\begin{itemize}\itemsep0pt}
\newcommand{\itemend}{\end{itemize}}
\newcommand{\descrstart}{\begin{description}\itemsep0pt}
\newcommand{\descrend}{\end{description}}
\newcommand{\enumstart}{\begin{enumerate}\itemsep0pt}
\newcommand{\enumend}{\end{enumerate}}

\newcommand{\ii}{\mathsf{i}}
\newcommand{\ee}{\mathsf{e}}

\usepackage{xspace}

\newcommand{\ignore}[1]{}

\def \s[#1]{\left(#1\right)}
\def \sA[#1]{\bigl(#1\bigr)}
\def \sB[#1]{\Bigl(#1\Bigr)}
\def \sC[#1]{\biggl(#1\biggr)}
\def \sD[#1]{\Biggl(#1\Biggr)}

\def \sk[#1]{\left[#1\right]}
\def \skA[#1]{\bigl[#1\bigr]}
\def \skB[#1]{\Bigl[#1\Bigr]}
\def \skC[#1]{\biggl[#1\biggr]}
\def \skD[#1]{\Biggl[#1\Biggr]}

\def \abs|#1|{\left| #1\right|}
\def \absA|#1|{\bigl|#1\bigr|}
\def \absB|#1|{\Bigl|#1\Bigr|}
\def \absC|#1|{\biggl|#1\biggr|}
\def \absD|#1|{\Biggl|#1\Biggr|}

\def \norm|#1|{\left\| #1\right\|}
\def \normA|#1|{\bigl\| #1\bigr\|}
\def \normB|#1|{\Bigl\| #1\Bigr\|}
\def \normC|#1|{\biggl\| #1\biggr\|}
\def \normD|#1|{\Biggl\| #1\Biggr\|}
\def \normS|#1|{\| #1\|}

\def \normFrob|#1|{\norm|#1|_{\mathrm{F}}}

\def \elem[#1]{[\![#1]\!]}

\def \ket#1|#2>%
{\ifx&#1&
|#2\rangle
\else
|#2\rangle_{\mathsf #1}
\fi}

\def \ketA#1|#2>%
{\ifx&#1&
\bigl|#2\bigr\rangle
\else
\bigl|#2\bigr\rangle_{\mathsf #1}
\fi}

\def \ip<#1,#2>{\langle #1, #2\rangle}
\def \ipA<#1,#2>{\bigl\langle #1, #2\bigr\rangle}
\def \ipB<#1,#2>{\Bigl\langle #1, #2\Bigr\rangle}
\def \ipC<#1,#2>{\biggl\langle #1, #2\biggr\rangle}
\def \ipD<#1,#2>{\Biggl\langle #1, #2\Biggr\rangle}

\newcommand{\polylog}{\mathop{\mathrm{polylog}}}

\usepackage{silence}
\def\mycommand#1#2{%
\draft{\marginpar{\fbox{\sf{#1:} $#2$}}}%
\expandafter\newcommand \csname#1\endcsname {#2}%
}

\smallone{\usepackage{algorithm}}
\bigone{\usepackage[chapter]{algorithm}}

\newcounter{asdf}
\newcounter{nomer}
\newlength{\otstup}
\newlength{\ostatok}
\newlength{\nomerwidth}
\newcommand{\algbegin}{\setcounter{asdf}{0}\setcounter{nomer}{0}}
\newcommand{\algend}{}

\release{}

\newcommand{\state}[1]{\refstepcounter{nomer}%
\makebox[\nomerwidth][r]{\footnotesize\arabic{nomer}:}\hspace{\value{asdf}\otstup}\hspace{1mm}%
\ostatok = \textwidth%
\addtolength{\ostatok}{-\value{asdf}\otstup}%
\addtolength{\ostatok}{-2\nomerwidth}%
\parbox[t]{\ostatok}{#1}}

\newcommand{\tab}{\addtocounter{asdf}{1}}
\newcommand{\untab}{\addtocounter{asdf}{-1}}

\usepackage{fullpage}
\release{\usepackage{hyperref}}

\title{Quantum Walks and Electric Networks}
\author{Aleksandrs Belovs\thanks{Faculty of Computing, University of Latvia, stiboh@gmail.com.}}
\date{}

\begin{document}
\maketitle

\begin{abstract}
We prove that a quantum walk can detect the presence of a marked element in a graph in $O(\sqrt{WR})$ steps for any initial probability distribution on vertices.  Here, $W$ is the total weight of the graph, and $R$ is the effective resistance.  This generalizes the result by Szegedy~\cite{szegedy:walk} that is only applicable if the initial distribution is stationary.  We describe a time-efficient quantum algorithm for 3-distinctness based on these ideas.
\end{abstract}

\section{Introduction}
Quantum walks are of great importance in quantum algorithms.  
For instance, quantum walk on the Johnson graph has been used in many applications: $k$-distinctness and $k$-sum problems~\cite{ambainis:distinctness, childs:subsetFinding}, triangle detection in a graph given by its adjacency matrix~\cite{magniez:triangle, jeffery:nestedWalks}, matrix product verification~\cite{buhrman:productVerification}, restricted range associativity~\cite{dorn:associativity}, and others.
For more examples, refer to the survey papers by Ambainis~\cite{ambainis:walkApplications},  Kempe~\cite{kempe:walkOverview}, or Santha~\cite{santha:walkBasedAlgorithms}.  

In this paper, we deal with the problem of detecting marked vertices in a graph.
Two main paradigms for this task, Szegedy-type quantum walks~\cite{szegedy:walk} and MNRS quantum walks~\cite{magniez:walkSearch}, assume the walk is started in the stationary distribution.  
In particular, Szegedy showed that the presence of a marked vertex can be determined in $O(\sqrt{H})$ steps of the quantum walk where $H$ is the average classical hitting time.   But starting in the stationary distribution can be a strong limitation if the graph is complex or not given in advance.
The main result of this paper is the generalization of the algorithm by Szegedy to arbitrary initial distribution.  In order to do so, we add two new ingredients to the analysis of Szegedy-type quantum walks:
\descrstart
\item[Electric Networks.]  A point of view on a graph as an electric network has turned out very fruitful in the analysis of classical random walks~\cite{doyle:walksElectric, bollobas:modernGraph}.  
But it seems to be completely ignored in the analysis of quantum walks.  The preceding papers relied on the spectral properties of the graph.

\item[Effective Spectral Gap Lemma. ] The effective spectral gap lemma is a very simple and powerful tool in the analysis of spectral properties of a special kind of unitary transformations~\cite{lee:stateConversion}.  This lemma has been used to prove the optimality of the adversary bound for quantum state conversion~\cite{lee:stateConversion}, and for the implementation of span programs as quantum algorithms~\cite{belovs:learningClaws}.  We show that the lemma can be also applied for general quantum walks.
\descrend

We show two examples of application of this quantum walk.  In \refsec{learning}, we show how a general learning graph~\cite{belovs:learning} can be implemented as a quantum walk.  In \refsec{kdist}, we use these ideas in a time-efficient quantum algorithm for $3$-distinctness.  The last example is interesting as a quantum walk on a graph not given in advance.  This is at the very heart of classical random walks: Since only local information is required to implement a random walk, they are often used to traverse graphs whose global structure is unknown (see, e.g.,~\cite{aleliunas:connectivity, schoning:ksat}).  Quantum walks require more global information than the classical ones, and they are usually used for graphs known in advance like for the Johnson graph mentioned above.

We hope the ideas presented in this paper will be useful for implementing other quantum walks.  Possible candidates could be time-efficient implementations of learning graphs and quantum analogues of random-walk-based algorithms.

The remaining part of the paper is organized as follows.  In \rf(sec:prelim), we recall the relations between classical hitting time and electric resistance of a graph, and some tools from quantum algorithms.  In \rf(sec:walk), we prove the main result, and in \rf(sec:learning), give an application to learning graphs.  In \rf(sec:kdist), we apply the new quantum walk algorithm for the 3-distinctness problem.

\section{Preliminaries}
\label{sec:prelim}
\subsection{Random Walks and Electric Networks}
\label{sec:circuits}
\mycommand{st}{\sigma}

Let $G=(V,E)$ be a simple undirected graph with each edge assigned a {\em weight} $w_e\ge 0$.  Let $W = \sum_{e\in E} w_e$ be the {\em total weight}.  Consider the following {\em random walk} on $G$:  If the walk is at a vertex $u\in V$, proceed to a vertex $v$ with probability proportional to $w_{uv}$, i.e., $w_{uv}/(\sum_{ux\in E} w_{ux})$.  The random walk has a {\em stationary probability distribution} $\pi=(\pi_u)$ given by $\pi_u = \sum_{uv\in E} w_{uv}/(2W)$.  One step of the random walk leaves $\pi$ unchanged.

Let $\st=(\st_u)$ be an {\em initial probability distribution} on the vertices of the graph, and let $M\subseteq V$ be some set of {\em marked vertices}.  We are interested in the {\em hitting time} $H_{\st, M}$ of the random walk: the expected number of steps of the random walk required to reach a vertex in $M$ when the initial vertex is sampled accordingly to $\st$.  If $\st$ is concentrated in a vertex $s\in V$, or $M$ consists of a single element $t\in V$, we often replace $\st$ by $s$ or $M$ by $t$.  For instance, we have $H_{\st, M} = \sum_{u\in V} \st_u H_{u, M}$.
We usually assume that $G$ and $\st$ are known, and the task is to determine whether $M$ is non-empty by performing the random walk.

Assume $M$ is non-empty, and define a {\em flow} on $G$ from $\st$ to $M$ as a real-valued function $p_e$ on the (oriented) edges of the graph satisfying the following conditions.  At first, $p_{uv} = -p_{vu}$.  Next, for each non-marked vertex $u$, the flow satisfies
\begin{equation}
\label{eqn:flowCondition}
\st_u = \sum_{uv\in E} p_{uv}.
\end{equation}
That is, $\st_u$ units of the flow are injected into $u$, it traverses through the graph, and is removed in a marked vertex.  Define the {\em energy} of the flow as 
\begin{equation}
\label{eqn:flowEnergy}
\sum_{e\in E} \frac{p_e^2}{w_e}.
\end{equation}
Clearly, the value of~\rf(eqn:flowEnergy) does not depend on the orientation of each $e$.
The {\em effective resistance} $R_{\st, M}$ is the minimal energy of
a flow from $\st$ to $M$.   For $R$, as for $H$, we also replace $\st$ and $M$ by the corresponding singletons.  
The resistance $R_{\st, M}$ equals the energy dissipated by the electric flow where the edges have conductance $w_e$, $\st_u$ units of the current are injected into each $u$, and then collected in $M$~\cite{bollobas:modernGraph}.  The following two results can be easily obtained from the results in Ref.~\cite{chandra:electrical}:


\begin{thm}
\label{thm:classical}
If $G$, $w$, $W$ are as above, $s$, $t$ are two vertices of $G$, $M\subseteq V$, and $\pi$ is the stationary distribution on $G$, then
\itemstart
\item[(a)] the commute time between $s$ and $t$, $H_{s,t} + H_{t,s}$, equals $2WR_{s,t}$;
\item[(b)] the average hitting time $H_{\pi, M}$ equals $2WR_{\pi, M}$.
\itemend
\end{thm}


We show that we obtain a quadratic improvement in the quantum case:  If $G$ and $\st$ are known in advance and the superposition $\sum_{u\in V} \sqrt{\st_u}\ket|u>$ is given, the presence of a marked vertex in $G$ can be determined in  $O(\sqrt{WR})$ steps of the quantum walk.
By combining this result with the second statement of \refthm{classical}, we obtain the main result of the paper by Szegedy~\cite{szegedy:walk}.

\subsection{Tools from Quantum Computing}
\label{sec:tools}
We assume the reader is familiar with the basics of quantum computation~\cite{chuang:quantum} and query complexity~\cite{buhrman:querySurvey}.  
Although we use the language of electric networks to state our results, our algorithms still use spectral properties of unitary transformations.  We start with a result we use to prove the existence of a spectral gap, and then we review how to detect it.


\begin{lem}[Effective Spectral Gap Lemma~\cite{lee:stateConversion}] \label{lem:effective}
Let $\Pi_A$ and $\Pi_B$ be two orthogonal projectors in the same vector space, and $R_A = 2\Pi_A-I$ and $R_B = 2\Pi_B-I$ be the reflections about their images.
Assume $P_\Theta$, where $\Theta\ge0$, is the orthogonal projector on the span of the eigenvectors of $R_BR_A$ with eigenvalues $\ee^{\ii\theta}$ such that $|\theta|\le \Theta$.  Then, for any vector $w$ in the kernel of $\Pi_A$, we have
\[ \|P_\Theta \Pi_B w \|\le \frac{\Theta}{2}\|w\|. \]
\end{lem}


\begin{thm}[Phase Estimation~\cite{kitaev:phaseEstimation, cleve:phaseEstimation}]
\label{thm:estimation}
Assume a unitary $U$ is given as a black box.  There exists a quantum algorithm that, given an eigenvector $\psi$ of $U$ with eigenvalue $\ee^{\ii\phi}$, outputs a real number $w$ such that $|w-\phi|\le\delta$ with probability at least $9/10$.  Moreover, the algorithm uses $O(1/\delta)$ controlled applications of $U$ and $\frac{1}{\delta}\polylog(1/\delta)$ other elementary operations.
\end{thm}

\section{Quantum Walk}
\label{sec:walk}
In this section, we construct a quantum counterpart of the random walk in \refsec{circuits}.  The quantum walk differs slightly from the quantum walk by Szegedy.  The framework of the algorithm goes back to~\cite{ambainis:formulaeEvaluation}, and \reflem{effective} is used to analyse its complexity.  We assume the notations of \refsec{circuits} throughout the section.

It is customary to consider quantum walks on bipartite graphs.  We keep with this tradition, and assume the graph $G=(V,E)$ is bipartite with parts $A$ and $B$.  Also, we assume the support of $\st$ is contained in $A$, i.e., $\st_u=0$ for all $u\in B$.  These are not very restrictive assumptions:  If either of them fails, consider the bipartite graph $G'$ with the vertex set $V' = V\times\{0,1\}$, the edge set $E' = \{(u,0)(v,1), (u,1)(v,0) \mid uv\in E\}$, edge weights $w'_{(u,0)(v,1)} =  w'_{(u,1)(v,0)} = w_{uv}$, the initial distribution $\st'_{(u,0)} = \st_u$, and the set of marked vertices $M' = M \times\{0,1\}$.  Then, for the new graph, $W' = 2W$, and $R'_{\st', M'} \le R_{\st, M}$.

\mycommand{qst}{\varsigma}

We assume the quantum walk starts in the state $\qst = \sum_{u\in V} \sqrt{\st_u}\;\ket|u>$ that is known in advance.  Also, we assume there is an upper bound $R$ known on the effective resistance from $\st$ to $M$ for all possible sets $M$ of marked states that might appear.

Now we define the vector space of the quantum walk.  
Let $S$ be the {\em support} of $\st$, i.e., the set of vertices $u$ such that $\st_u\ne 0$.  
The vectors $\{\ket|u> \mid u\in S\}\cup\{\ket|e>\mid e\in E\}$ form the computational basis of the vector space of the quantum walk.  Let $\cH_u$ denote the {\em local space} of $u$, i.e., the space spanned by $\ket|uv>$ for $uv\in E$ and $\ket|u>$ if $u\in S$.  We have that $\bigoplus_{u\in A} \cH_u$ equals the whole space of the quantum walk, and $\bigoplus_{u\in B} \cH_u$ equals the subspace spanned by the vectors $\ket|e>$ for $e\in E$.

The {\em step of the quantum walk} is defined as $R_BR_A$ where $R_A = \bigoplus_{u\in A} D_u$ and $R_B = \bigoplus_{u\in B} D_u$ are the direct sums of the {\em diffusion} operations.  Each $D_u$ is a reflection operation in $\cH_u$.  Hence, all $D_u$ in $R_A$ or $R_B$ commute, that makes them easy to implement in parallel.  They are as follows:
\itemstart
\item If a vertex $u$ is marked, then $D_u$ is the identity, i.e., the reflection about $\cH_u$;
\item If $u$ is not marked, then $D_u$ is the reflection about the orthogonal complement of $\psi_u$ in $\cH_u$, where
\begin{equation}
\label{eqn:psi}
\psi_u = \sqrt{\frac{\st_u}{C_1R}}\; \ket|u> + \sum_{uv\in E} \sqrt{w_{uv}}\;\ket|uv>
\end{equation}
for some constant $C_1>0$.  This also holds for $u\notin S$: For them, the first term in~\rf(eqn:psi) disappears.
\itemend

\begin{algorithm}
\caption{The quantum walk algorithm.  Here, $C$ is some constant to be specified later.}
\label{alg:walk}
\algbegin
\state{Start in the state $\qst$.}
\state{Calculate, for each $u\in S$, whether it is marked, and measure this bit.}
\state{{\bf If} the result of the measurement shows `marked', }
\tab
\state{{\bf then} output ``marked vertices exist'', and {\bf quit}.}
\untab
\state{Execute quantum phase estimation on $R_BR_A$ with precision $1/(C\sqrt{RW})$.}
\state{{\bf If} the eigenvalue is 1, }
\tab
\state{{\bf then} output ``marked vertices exist'';}
\state{{\bf otherwise}, output ``no marked vertices''.}
\algend
\end{algorithm}


\begin{thm}
\label{thm:walk}
\refalg{walk} detects the presence of a marked vertex with probability at least $2/3$.  The algorithm uses $O(\sqrt{RW})$ steps of the quantum walk.
\end{thm}

\pfstart
The second statement follows immediately from \refthm{estimation}.  Let us prove the correctness.
If a vertex in the initial distribution is marked with probability at least $2/3$, then this is detected at Step 3 with the same probability, and we are done.  So, assume this probability is less than $2/3$, and the measurement outcome on Step 3 is `not marked'.  Then, the state of the algorithm collapses to a state $\qst'$ with the support disjoint from $M$, and $R_{\qst', M}\le 9R$.  Thus, we further assume $S$ is disjoint from $M$.  

Let us consider Steps 5---8 of the algorithm.  We start with the positive case.  Let $p_e$ be a flow from $\st$ to $M$ with energy at most $R$. At first, using the Cauchy-Schwarz inequality and that $S$ is disjoint from $M$, we get
\begin{equation}
\label{eqn:RW}
RW \ge \sC[\sum_{e\in E} \frac{p_e^2}{w_e}]\sC[\sum_{e\in E} w_e] \ge \sum_{e\in E} |p_e| \ge 1.
\end{equation}
Now, we construct an eigenvalue-1 eigenvector
\[
\phi = \sqrt{C_1R} \sum_{u\in S} \sqrt{\st_u} \ket|u> - \sum_{e\in E} \frac{p_e}{\sqrt{w_e}}\ket|e>
\]
of $R_BR_A$ having large overlap with $\qst$  (assume the orientation of each edge $e$ is from $A$ to $B$.)
Indeed, by~\refeqn{flowCondition}, $\phi$ is orthogonal to all $\psi_u$, hence, is invariant under the action of both $R_A$ and $R_B$.  Moreover, $\|\phi\|^2 = C_1R + \sum_{e\in E} p_e^2/w_e$, and $\ip<\phi,\qst> = \sqrt{C_1R}$.  Since we assumed $R\ge \sum_{e\in E} p_e^2/w_e$, we get that the normalized vector satisfies 
\begin{equation}
\label{eqn:positive}
\ipB<\frac\phi{\|\phi\|},\;\qst>\ge \sqrt{\frac{C_1}{1+C_1}}.
\end{equation}
Now consider the negative case.  Let $w$ be defined by
\[
w = \sqrt{C_1R}\sC[\sum_{u\in S} \sqrt{\frac{\st_u}{C_1R}}\; \ket|u> + \sum_{e\in E} \sqrt{w_e}\;\ket|e>].
\]
Let $\Pi_A$ and $\Pi_B$ be the projectors on the invariant subspaces of $R_A$ and $R_B$, respectively.  Since $S\subseteq A$, we get that $\Pi_Aw=0$ and $\Pi_Bw = \qst$.  Hence, by \reflem{effective}, we have that, if
\[
\Theta = \frac{1}{C_2\sqrt{1+C_1 RW}}
\]
for some constant $C_2>0$, then the overlap of $\qst$ with the eigenvectors of $R_BR_A$ with phase less than $\Theta$ is at most $1/(2C_2)$.  Comparing this with~\refeqn{positive}, we get that it is enough to execute phase estimation with precision $\Theta$ if $C_1$ and $C_2$ are large enough.  Also, assuming $C_1\ge 1$, we get $\Theta = \Omega(1/\sqrt{RW})$ by~\refeqn{RW}.
\pfend

\section{Application: Learning Graphs}
\newcommand{\reg}[1]{\mathsf{#1}}
\mycommand{btr}{\bigtriangleup}
\label{sec:learning}
As a simple example, we consider an alternative way of implementing learning graphs~\cite{belovs:learning}.  Originally, learning graphs were implemented via the dual adversary bound~\cite{lee:learningKdistPrior} that is later transformed into a quantum walk~\cite{reichardt:advTight, lee:stateConversion}.  Although this gives a query-efficient implementation, the time-efficiency of this approach is rather unsatisfactory.  

Hence, from the perspective of time-efficiency, it would be preferable to implement a learning graph as a quantum walk directly.  There was one attempt of doing so in Ref.~\cite{jeffery:nestedWalks}.  The authors use an MNRS-type quantum walk~\cite{magniez:walkSearch}, and give quantum walks corresponding to a number of previously developed learning graphs.  We, however, use the Szegedy-type quantum walk, and our construction is valid for an arbitrary learning graph.

A learning graph can be defined as a special case of a quantum walk from \refsec{walk}.  A learning graph computes a function $f\colon\cD\to\{0,1\}$ with $\cD\subseteq [q]^n$.  Vertices of the graph are subsets of $[n]$, and the allowed edges are only between vertices $S$ and $S\cup\{j\}$ for some $S\subset [n]$ and $j\in[n]\setminus S$.  
The initial distribution $\st$ is concentrated on the vertex $\emptyset$.  For each positive input $x\in f^{-1}(1)$, a vertex $S$ is marked if and only if it contains a 1-certificate for $x$, i.e., $f(z) = 1$ for all $z\in\cD$ such that $z_S = x_S$.  The {\em complexity} of the learning graph is defined as $\sqrt{WR}$ in the notations of \refsec{circuits}.  It is known~\cite{lee:learningKdistPrior} that then the quantum query complexity of $f$ is $O(\sqrt{WR})$.  We show this again using a quantum walk from \rf(sec:walk).  

The learning graph is a bipartite graph: the part $A$ contains all vertices of even cardinality, and the part $B$ contains all vertices of odd cardinality.  Also, the support of $\st$ is concentrated in $A$.  Hence, the algorithm from \refsec{walk} can be applied, and the presence of a marked vertex can be detected in $O(\sqrt{WR})$ steps of the quantum walk.  It suffices to show that one step of the quantum walk can be implemented in $O(1)$ quantum queries.

This can be done using standard techniques.  Let $x$ be an input to $f$ given as an oracle.  The quantum walk has two registers: the {\em data register} $\reg D$, and the {\em coin register} $\reg C$.  The first register can be in a state $\ket D|S>$ for some $S\subseteq[n]$.  The register contains the description of the subset $S$, the values of $x_j$ for $j\in S$, and some ancillary information.  Because of the interference, it is important that $\ket D|S>$ is always represented in exactly the same way that only depends on $S$ and the input string $x$.  The second register stores an element $j\in [n]$.  An element $\ket D|S>\ket C|j>$ of the computational basis represents the edge of the learning graph connecting subsets $S$ and $S\btr\{j\}$, where $\btr$ is the symmetric difference.  Additionally, there is the state $\ket D|\emptyset>$ for the initial distribution.

The step of the quantum walk is performed as follows.  Start with a superposition of $\ket|\emptyset>$ and the states of the form $\ket D|S>\ket C|j>$ with $S$ in $A$.  At first, perform the reflection $R_A$ as described in \refsec{walk}.  It is possible to detect whether $S$ is marked by considering the values $x_j$ stored in $\ket D|S>$, and $\psi_S$ does not depend on the input.  Hence, this operation does not require any oracle queries.  Next, apply the {\em update operation} that maps $\ket D|S>\ket C|j>$ into $\ket D|S\btr \{j\}>\ket C|j>$.  This represents the same edge, but with the content of the data register in $B$.  The update operation requires one oracle query in order to compute or uncompute $x_j$.  After that, perform $R_B$ similarly to $R_A$, and apply the update operation once more.  Hence, one step of the quantum walk requires $O(1)$ oracle queries, and $f(x)$ can be computed in $O(\sqrt{WR})$ quantum queries.

\section{Application: 3-distinctness}
\mycommand{tO}{\tilde O}
\label{sec:kdist}
In \refsec{learning}, we demonstrated that the quantum walk algorithm from \refsec{walk} can be used to implement learning graphs.  In this section, we show an application that uses techniques unavailable for ordinary learning graphs.  

Consider the $k$-distinctness problem (the definition follows shortly).  The first quantum algorithm for this problem was constructed by Ambainis~\cite{ambainis:distinctness} using quantum walk on the Johnson graph.  This requires $O(n^{k/(k+1)})$ queries and can be implemented in the same time complexity up to polylogarithmic factors.  For $k=2$, it is tight~\cite{shi:collisionLower}.

Recently, the query complexity of the problem was improved to $o(n^{3/4})$ using a generalized learning graph approach~\cite{belovs:learningKDist}.  For $k=3$, it gives $O(n^{5/7})$ queries. 
However, it is unknown how to implement the algorithm time-efficiently.  
In this section, we describe a quantum algorithm for 3-distinctness having the same time complexity up to polylogarithmic factors.  This is a different algorithm from Ref.~\cite{belovs:learningKDist}, and is based on ideas from Ref.~\cite{lee:learningKdistPrior}.  Formally, we prove the following result.

\begin{thm}
\label{thm:3dist}
The 3-distinctness problem can be solved by a quantum algorithm in time $\tO(n^{5/7})$ using quantum random access quantum memory (QRAQM) of size $\tO(n^{5/7})$.
\end{thm}

Recall that the Ambainis' algorithm consists of two phases: the set-up phase that prepares the uniform superposition, and the quantum walk itself.  Our algorithm also consists of these two phases.  Moreover, for $k=2$, it is exactly the Ambainis' algorithm.  Interestingly, in our case, the analysis of the quantum walk is quite simple, and can be easily generalized to any $k$.  It is the set-up phase that is hard to generalize.  The case of $k=3$ has a relatively simple {\em ad hoc} solution that we describe in \rf(sec:setup).

During the preparation of the paper, we learned about an alternative time-efficient quantum algorithm for the 3-distinctness problem by Andrew Childs, Stacey Jeffery, Robin Kothari and Fr\'ed\'eric Magniez (personal communication).  They use an MNRS-type quantum walk.  Their algorithm has a similar set-up phase as ours, but a more complicated quantum walk phase, that is hard to generalize to arbitrary $k$.

\subsection{Technicalities}
\label{sec:3disttech}
We start the section with some notations and algorithmic primitives we need for our algorithm.  For more detail on the implementation of these primitives, refer to the paper by Ambainis~\cite{ambainis:distinctness}.  Although this paper does not exactly give the primitives we need, it is straight-forward to apply the necessary modifications, so we don't go into the detail.

Some parts of our algorithm work for the general $k$-distinctness problem, so we describe the notation for this problem.  We are given a string $x\in [q]^n$.  A subset $J\subseteq[n]$ of size $\ell$ is called an {\em $\ell$-collision} iff $x_i=x_j$ for all $i,j\in J$.  In the $k$-distinctness problem, the task is to determine whether the given input contains a $k$-collision.  Inputs with a $k$-collision are called {\em positives}, the remaining ones are called {\em negative}.

Without loss of generality, we may assume that any positive input contains exactly one $k$-collision.  Otherwise, one can first try random subinstances of the problem, and this reduction can be made time-efficient~\cite{ambainis:distinctness}.  Also, we may assume there are $\Omega(n)$ $(k-1)$-collisions by extending the input with dummy elements.

For a subset $S\subseteq[n]$ and $i\in[k]$, let $S_i$ denote the set of $j\in S$ such that $|\{j'\in S \mid x_{j'} = x_j\}| = i$.  Denote $r_i = |S_i|/i$, and call $\tau = (r_1,\dots,r_k)$ the {\em type} of $S$.


Our main technical tool is a dynamical quantum data structure that maintains a subset $S\subseteq[n]$ and the values $x_j$ for $j\in S$.  We use notation $\ket D|S>$ to denote a register containing the data structure for a particular choice of $S\subseteq [n]$.

The data structure is capable of performing a number of operations in polylogarithmic time.  The initial state of the data structure is $\ket D|\emptyset>$.  The update operation adds or removes an element:  $\ket D|S>\ket |j>\ket |x_j>\mapsto \ket D|S\btr \{j\}> \ket |j>\ket |0>$.  Recall that $\btr$ stands for the symmetric difference.
There is a number of query operations to the data structure.  It is able to give the type $\tau$ of $S$.  For integers $i\in[k]$ and $\ell\in [|S_i|]$, it returns the $\ell$th element of $S_i$ according to some internal ordering.  Given an element $j\in[n]$, it detects whether it is in $S$, and if it is, returns the tuple $(i,\ell)$ such that $j$ is the $\ell$th element of $S_i$.  Given $a\in[q]$, it returns $i\in[k]$ such that $a$ equals to a value in $S_i$ or says there is no such $i$.

The data structure is coherence-friendly, i.e., a subset $S$ will have the same representation $\ket D|S>$ independently of the sequence of update operations that results in this subset.  Next, it has an exponentially small error probability of failing that can be ignored.  Finally, the implementation of this data structure requires quantum random access quantum memory (QRAQM) in the terms of Ref.~\cite{kuperberg:anotherDihedral}.

\subsection{Quantum Walk}
\label{sec:3distwalk}
In this section, we describe the quantum walk part of the algorithm.  Formally, it is as follows.

\begin{lem}
\label{lem:kdist}
Let $r_1,\dots,r_{k-1} = o(n)$ be positive integers, $x\in[q]^n$ be an input for the $k$-distinctness problem, and $V_0$ be the set of $S\subseteq[n]$ having type $(r_1,\dots,r_{k-1},0)$.
Given the uniform superposition $\qst = \frac{1}{\sqrt{|V_0|}} \sum_{S\in V_0} \ket|S>$, it is possible to solve the $k$-distinctness problem in $\tO(n/\sqrt{\min\{r_1,\dots,r_{k-1}\}})$ quantum time.
\end{lem}

\pfstart
As mentioned in \rf(sec:3disttech), we may assume that any input contains at most one $k$-collision and $\Omega(n)$ $(k-1)$-collisions.  Define $r_k=0$, and the type $\tau_i$ as $(r_1,\dots,r_{i-1}, r_i+1, r_{i+1},\dots,r_k)$ for $i\in[0,k]$.  Let $V_i$ be the set of all $S\subseteq[n]$ having type $\tau_i$.  It is consistent with our previous notation for $V_0$.  Denote $V=\bigcup_i V_i$.  Also, for $i\in[k]$, define the set $Z_i$ of {\em dead-ends}  consisting of vertices of the form $(S,j)$ for $S\in V_{i-1}$ and $j\in [n]$ such that $S\btr \{j\}\notin V$.  Again, $Z=\bigcup_i Z_i$.

The vertex set of $G$ is $V\cup Z$.  Each $S\in V\setminus V_k$ is connected to $n$ vertices: one for each $j\in [n]$.  If $S\btr\{j\}\in V$, it is the vertex $S\btr\{j\}$, otherwise, it is $(S,j)\in Z$.  A vertex $S\in V_k$ is connected to $k$ vertices in $V_{k-1}$ differing from $S$ in one element.   Each $(S,j)\in Z$ is only connected to $S$.  
The weight of each edge is 1.
A vertex is marked if and only if it is contained in $V_k$.

\refalg{walk} is not directly applicable here because we do not know the graph in advance (it depends on the input), nor we know the amplitudes in the initial distribution $\qst$.  However, we know the graph locally, and our ignorance in the amplitudes of $\qst$ conveniently cancels out with our ignorance in the size of $G$.

Let us briefly describe the implementation of the quantum walk on $G$ following \refsec{walk}.  Let $G=(V\cup Z,E)$ be the graph described above. It is bipartite: The part $A$ contains all $V_i$ and $Z_i$ for $i$ even, and $B$ contains all $V_i$ and $Z_i$ for $i$ odd.  The support of $\qst$ is contained in $A$.  The reflections $R_A$ and $R_B$ are the direct sums of local reflections $D_u$ over all $u$ in $A$ and $B$, respectively.  They are as follows:
\itemstart
\item If $u\in V_k$, then $D_u$ is the identity in $\cH_u$.
\item If $u\in Z_i$, then $D_u$ negates the amplitude of the only edge incident to $u$.
\item If $u\in V_i$ for $i<k$, then $D_u$ is the reflection about the orthogonal complement of $\psi_u$ in $\cH_u$.  If $u\in V_0$, or $u\in V_i$ with $i>0$, then $\psi_u$ is defined as
\[
\psi_u = \frac{1}{\sqrt{C_1}} \ket |u> + \sum_{uv\in E} \ket|uv>,\qquad\text{or}\qquad
\psi_u = \sum_{uv\in E} \ket|uv>,
\]
respectively.  Here, $C_1$ is a constant.
\itemend

The space of the algorithm consists of three registers: $\reg D$, $\reg C$ and $\reg Z$.  The data register $\reg D$ contains the data structure for $S\subseteq[n]$.  The coin register $\reg C$ contains an integer in $[0,n]$, and the qubit $\reg Z$ indicates whether the vertex is an element of $Z$.  A combination $\ket D|S>\ket C|0>\ket Z|0>$ with $S\in V_0$ indicates a vertex in $V_0$ that is used in $\qst$.  A combination $\ket D|S> \ket C|j>\ket Z|0>$ with $j>0$ indicates the edge between $S$ and $S\btr \{j\}$ or $(S,j)\in Z$.  Finally, a combination $\ket D|S>\ket C|j>\ket Z|1>$ indicates the edge between $(S,j)\in Z$ and $S\in V$.

Similarly to \rf(sec:learning), the reflections $R_A$ and $R_B$ are broken down into the diffuse and update operations.  The diffuse operations perform the local reflections in the list above. 
For the first one, do nothing conditioned on $\ket D|S>$ being marked.  For the second one, negate the phase conditioned on $\reg Z$ containing 1.  The third reflection is the Grover diffusion~\cite{grover:search} with one special element if $S\in V_0$.  Similarly to \rf(alg:walk), the orientation of the edges may be ignored because the graph is bipartite.

The update operation can be performed using the primitives from \rf(sec:3disttech).  Given $\ket D|S> \ket C|j>\ket Z|b>$, calculate whether $S\btr\{j\}\in V$ in a fresh qubit $\reg Y$.  Conditioned on $\reg Y$, query the value of $x_j$ and perform the update operation for the data structure.  Conditioned on $\reg Y$ not being set, flip the value of $\reg Z$.  Finally, uncompute the value in $\reg Y$.  On the last step, we use that $\ket D|S>\ket C|j>$ represents an edge between vertices in $V$ if and only if $\ket D|S\btr\{j\}> \ket C|j>$ does the same.

After we showed how to implement the step of the quantum walk efficiently, let us estimate the required number of steps.  The argument is very similar to the one in \rf(thm:walk).  Let us start with the positive case.  Assume $\{a_1,\dots,a_k\}$ is the unique $k$-collision.
Let $V_0'$ denote the set of $S\in V_0$ that are disjoint from $\{a_1,\dots,a_k\}$, and $\st'$ be the uniform probability distribution on $V_0'$.  
Define the flow $p$ from $\st'$ to $V_k$ as follows.  For each $S\in V_i$ such that $i<k$ and $S\cap M = \{a_1,\dots,a_i\}$, define flow $p_e = 1/|V_0'|$ on the edge $e$ from $S$ to $S\cup\{a_{i+1}\}\in V_{i+1}$.  Define $p_e = 0$ for all other edges $e$.
Let
\[
\phi = \sqrt{C_1} \sum_{S\in V_0'}\frac{1}{|V_0'|} \ket|S> - \sum_{e\in E} p_e\ket|e>.
\]
This vector is orthogonal to all $\psi_u$, hence, is invariant under the action of $R_BR_A$.
Also, $\|\phi\|^2 = (k+C_1)/|V_0'|$, and $\ip<\phi, \qst> = \sqrt{C_1/|V_0|}$.  Hence,
\[
\ipB<\frac\phi{\|\phi\|},\;\qst> = \sqrt{\frac{C_1|V_0'|}{(k+C_1)|V_0|}} \sim \sqrt{\frac{C_1}{k+C_1}}
\]
where $\sim$ stands for the asymptotic equivalence as $n\to\infty$.

In the negative case, define
\[
w = \sqrt{\frac{C_1}{|V_0|}} 
\sC[\sum_{S\in V_0} \frac{1}{\sqrt{C_1}} \ket|S> + \sum_{e\in E} \ket|e>].
\]
Similarly to the proof of \refthm{walk}, we have that $\Pi_Aw=0$ and $\Pi_Bw = \qst$.

Let us estimate $\|w\|$.  The number of edges in $E$ is at most $n$ times the number of vertices in $V_0\cup\cdots\cup V_{k-1}$.  Thus, we have to estimate $|V_i|$ for $i\in[k-1]$.  Consider the relation between $V_0$ and $V_i$ where $S\in V_0$ and $S'\in V_i$ are in the relation iff $S'\setminus S$ consists of $i$ equal elements. Each element of $V_0$ has at most $n{k-1\choose i} = O(n)$ images in $V_i$ because there are at most $n$ maximal collisions in the input, and for each of them, there are at most ${k-1\choose i}$ variants to extend $S$ with.  On the other hand, each element in $V_i$ has exactly $r_i+1$ preimages in $V_0$.  Thus, $|V_i| =O( n |V_0|/r_i )$.  Thus,
\[
\|w\| = O\s[ \sqrt{1+n/r_1+n/r_2+\cdots+n/r_{k-1}} ] =
O\s[ {n}/{\sqrt{\min\{r_1,\dots, r_{k-1}\}}} ].
\]
By \reflem{effective}, we have that if
$\Theta = \Omega(1/\|w\|)$,
then the overlap of $\qst$ with the eigenvectors of $R_BR_A$ with phase less than $\Theta$ can be made at most $1/C_2$ for any constant $C_2>0$.  Thus, it is enough to execute the phase estimation with precision $\Theta$ if $C_1$ and $C_2$ are large enough.  By \rf(thm:estimation), this requires $O( {n}/{\sqrt{\min\{r_1,\dots, r_{k-1}\}}} )$ iterations of the quantum walk.
\pfend

\subsection{Preparation of the Initial State}
\label{sec:setup}
Now we describe how to generate the uniform superposition $\qst$ over all elements in $V_0$ from the formulation of \rf(lem:kdist) efficiently in the special case of $k=3$.  Let us denote $r_1 = n^{5/7}$ and $r_2 = n^{4/7}$.  We start in the assumption the input is negative.

Prepare the state ${n\choose r_1}^{-1/2} \sum_{S: |S| = r_1} \ket D|S>$ in time $\tO(r_1)$.  This is very similar to the algorithm by Ambainis, and we omit the details.  
  Measure the type of $S$.  
  The state of the algorithm collapses to the uniform superposition of the subsets of some type $\tau = (t_1,t_2)$.  Unfortunately, with high probability, $t_2$ will be of order $r_1^2/n$ that is much smaller than the required size $r_2$.

We enlarge the size of $S_2$ by using the Grover search repeatedly.  For each $S$ in the superposition, apply the Grover search over $[n]$.  An element $j\in[n]$ is marked iff $j\notin S$ and $x_j$ is equal to an element in $S_1$.  This can be tested using the primitives from \rf(sec:3disttech). If the Grover search fails, repeat it from the current state.  If the search succeeds, the state is a superpositon of states of the form $\ket D|S> \ket |j>$.  Query the value of $x_j$, and update the data structure.  This gives a superposition over $\ket D|S\cup\{j\}> \ket |j>$.  Let $S' = S\cup\{j\}$.  Apply the primitive that transforms $j$ into its number in $S'_2$.  This gives a superposition over $\ket D|S'>\ket |i>$ where $i\in [|S'_2|]$.  We show in a moment that, for a fixed $S'$, all states $\ket D|S'>\ket |i>$ have the same amplitude, hence, the second register can be detached.

A typical subset has $\Omega(r_1)$ elements in $S_1$ that can be extended to a 2-collision, hence, the Grover search requires $O(\sqrt{n/r_1})$ iterations.  As we load $O(r_2)$ additional elements, the time spent during the Grover search is $\tO(r_2\sqrt{n/r_1})$.

Now assume each $S$ contains $r_2$ 2-collisions.  Unfortunately, the state is not the uniform superposition we require for the quantum walk in \rf(lem:kdist).  But due to symmetry, at any place in the algorithm, the amplitude of a subset $S$ only depends on the number of elements in $S_1$ that can be extended to a 2-collision.  This shows that, indeed, the second register can be detached after the Grover search.  Moreover, this gives us a way to generate the uniform superposition we require.

We measure the content of $S_1$.  Let $B$ be the outcome.  The state collapses to the uniform superposition over subsets $S_2$ consisting of $r_2$ 2-collisions not using the values in $B$.  Then, we repeat the first step, i.e., for each $S$, we construct the uniform superposition over subsets of size $r_1$ consisting of elements outside $S_2$ and having values different from the ones in $B$.  After that, we measure the type of the subset.  This results in the uniform superposition over states in $V_0$ of type $(r_2', r_1')$ with $r_2'>r_2$ and $r_1'=\Theta(r_1)$ and avoiding elements with values in $B$.

In the positive case, due to a similar argument, the state can be written as $\alpha\qst' + \sqrt{1-\alpha^2}\qst''$ where $\qst'$ is the uniform superposition over $V_0'$ as defined in the proof of \rf(lem:kdist), and $\qst''$ is some superposition over $\ket D|S>$ where $S$ intersects $\{a_1,a_2,a_3\}$.  One can show that $\alpha$ is close to 1, hence, the initial state has large overlap with eigenvalue-1 eigenspace of $R_BR_A$.%
\footnote{One can modify the algorithm so that it does not require this observation.  With probability $1/2$, continue with the old algorithm, and with probability $1/2$, measure the content of $S$ and search for a 2-distinctness outside $S$ having a value equal to a value in $S$.  This can be done using the standard algorithm for 2-distinctness with minor modifications.}

Then, we can apply the algorithm from \reflem{kdist} with additional modification that a vertex $(S,j)$ is declared a dead-end also if $x_j$ has a value in $B$.  This finds a 3-collision in time $\tO(n/\sqrt{r_2})$ if its value is different from a value in $B$.  For the values in $B$, we search for a 2-collision outside $B$ but having a value equal to a value in $B$.  This can be implemented in time $\tO(n^{2/3})$ using the standard algorithm for 2-distinctness with minor modifications.

Thus, up to polylogarithmic factors, the time complexity of the algorithm is
\(
r_1 + r_2\sqrt{n/r_1} + n/\sqrt{r_2}.
\)
This attains optimal value of $\tO(n^{5/7})$ for $r_1 = n^{5/7}$ and $r_2 = n^{4/7}$.  This finishes the proof of \rf(thm:3dist).

\section*{Acknowledgments}
I would like to thank Andris Ambainis, Andrew Childs, Stacey Jeffery, Robin Kothari, Troy Lee, Fr\'ed\'eric Magniez and Miklos Santha for useful discussions and remarks about the presentation of the paper.

This work has been supported by the European Social Fund within the project ``Support for Doctoral Studies at University of Latvia'' and by FET-Open project QCS.

\release{\bibliographystyle{\relativepath habbrvE}}
\bibliography{../../bib}

\begin{thebibliography}{10}

\bibitem{shi:collisionLower}
S.~Aaronson and Y.~Shi.
\newblock Quantum lower bounds for the collision and the element distinctness
  problems.
\newblock {\em Journal of the ACM}, 51(4):595--605, 2004.

\bibitem{aleliunas:connectivity}
R.~Aleliunas, R.~M. Karp, R.~J. Lipton, L.~Lovasz, and C.~Rackoff.
\newblock Random walks, universal traversal sequences, and the complexity of
  maze problems.
\newblock In {\em Proc. of 20th IEEE FOCS}, pages 218--223, 1979.

\bibitem{ambainis:walkApplications}
A.~Ambainis.
\newblock Quantum walks and their algorithmic applications.
\newblock {\em International Journal of Quantum Information}, 1(4):507--518,
  2003,  \href{http://xxx.lanl.gov/abs/quant-ph/0403120}{{\ttfamily
  arXiv:quant-ph/0403120}}.

\bibitem{ambainis:distinctness}
A.~Ambainis.
\newblock Quantum walk algorithm for element distinctness.
\newblock {\em SIAM Journal on Computing}, 37(1):210--239, 2007,
  \href{http://xxx.lanl.gov/abs/quant-ph/0311001}{{\ttfamily
  arXiv:quant-ph/0311001}}.

\bibitem{ambainis:formulaeEvaluation}
A.~Ambainis, A.~M. Childs, B.~W. Reichardt, R.~{\v Spalek}, and S.~Zhang.
\newblock Any {AND-OR} formula of size {$N$} can be evaluated in time
  {$N^{1/2+o(1)}$} on a quantum computer.
\newblock {\em SIAM Journal on Computing}, 39(6):2513--2530, 2010.

\bibitem{belovs:learningKDist}
A.~Belovs.
\newblock Learning-graph-based quantum algorithm for $k$-distinctness.
\newblock In {\em Proc. of 53rd IEEE FOCS}, pages 207--216, 2012,
  \href{http://xxx.lanl.gov/abs/1205.1534}{{\ttfamily arXiv:1205.1534}}.

\bibitem{belovs:learning}
A.~Belovs.
\newblock Span programs for functions with constant-sized 1-certificates.
\newblock In {\em Proc. of 44th ACM STOC}, pages 77--84, 2012,
  \href{http://xxx.lanl.gov/abs/1105.4024}{{\ttfamily arXiv:1105.4024}}.

\bibitem{lee:learningKdistPrior}
A.~Belovs and T.~Lee.
\newblock Quantum algorithm for $k$-distinctness with prior knowledge on the
  input.
\newblock 2011,  \href{http://xxx.lanl.gov/abs/1108.3022}{{\ttfamily
  arXiv:1108.3022}}.

\bibitem{belovs:learningClaws}
A.~Belovs and B.~W. Reichardt.
\newblock Span programs and quantum algorithms for $st$-connectivity and claw
  detection.
\newblock In {\em Proc. of 20th ESA}, volume 7501 of {\em LNCS}, pages
  193--204, 2012,  \href{http://xxx.lanl.gov/abs/1203.2603}{{\ttfamily
  arXiv:1203.2603}}.

\bibitem{bollobas:modernGraph}
B.~{Bollob\'as}.
\newblock {\em Modern graph theory}, volume 184 of {\em Graduate Texts in
  Mathematics}.
\newblock Springer, 1998.

\bibitem{buhrman:querySurvey}
H.~Buhrman and R.~de~Wolf.
\newblock Complexity measures and decision tree complexity: a survey.
\newblock {\em Theoretical Computer Science}, 288:21--43, 2002.

\bibitem{buhrman:productVerification}
H.~Buhrman and R.~{\v Spalek}.
\newblock Quantum verification of matrix products.
\newblock In {\em Proc. of 17th ACM-SIAM SODA}, pages 880--889, 2006,
  \href{http://xxx.lanl.gov/abs/quant-ph/0409035}{{\ttfamily
  arXiv:quant-ph/0409035}}.

\bibitem{chandra:electrical}
A.~K. Chandra, P.~Raghavan, W.~L. Ruzzo, R.~Smolensky, and P.~Tiwari.
\newblock The electrical resistance of a graph captures its commute and cover
  times.
\newblock {\em Computational Complexity}, 6(4):312--340, 1996.

\bibitem{childs:subsetFinding}
A.~M. Childs and J.~M. Eisenberg.
\newblock Quantum algorithms for subset finding.
\newblock {\em Quantum Information \& Computation}, 5(7):593--604, 2005,
  \href{http://xxx.lanl.gov/abs/quant-ph/0311038}{{\ttfamily
  arXiv:quant-ph/0311038}}.

\bibitem{cleve:phaseEstimation}
R.~Cleve, A.~Ekert, C.~Macchiavello, and M.~Mosca.
\newblock Quantum algorithms revisited.
\newblock {\em Proceedings of the Royal Society of London A: Mathematical,
  Physical and Engineering Sciences}, 454(1969):339--354, 1998,
  \href{http://xxx.lanl.gov/abs/quant-ph/9708016}{{\ttfamily
  arXiv:quant-ph/9708016}}.

\bibitem{dorn:associativity}
S.~{D\"orn} and T.~Thierauf.
\newblock The quantum query complexity of algebraic properties.
\newblock In {\em Proc. of 16th FCT}, volume 4639 of {\em LNCS}, pages
  250--260. Springer, 2007,
  \href{http://xxx.lanl.gov/abs/0705.1446}{{\ttfamily arXiv:0705.1446}}.

\bibitem{doyle:walksElectric}
P.~G. Doyle and J.~L. Snell.
\newblock {\em Random Walks and Electric Networks}, volume~22 of {\em Carus
  Mathematical Monographs}.
\newblock MAA, 1984,  \href{http://xxx.lanl.gov/abs/math.PR/0001057}{{\ttfamily
  arXiv:math.PR/0001057}}.

\bibitem{grover:search}
L.~K. Grover.
\newblock A fast quantum mechanical algorithm for database search.
\newblock In {\em Proc. of 28th ACM STOC}, pages 212--219, 1996.

\bibitem{jeffery:nestedWalks}
S.~Jeffery, R.~Kothari, and F.~Magniez.
\newblock Nested quantum walks with quantum data structures.
\newblock 2012,  \href{http://xxx.lanl.gov/abs/1210.1199}{{\ttfamily
  arXiv:1210.1199}}.

\bibitem{kempe:walkOverview}
J.~Kempe.
\newblock Quantum random walks: an introductory overview.
\newblock {\em Contemporary Physics}, 44(4):307--327, 2003,
  \href{http://xxx.lanl.gov/abs/quant-ph/0303081}{{\ttfamily
  arXiv:quant-ph/0303081}}.

\bibitem{kitaev:phaseEstimation}
A.~Kitaev.
\newblock Quantum measurements and the abelian stabilizer problem.
\newblock 1995,  \href{http://xxx.lanl.gov/abs/quant-ph/9511026}{{\ttfamily
  arXiv:quant-ph/9511026}}.

\bibitem{kuperberg:anotherDihedral}
G.~Kuperberg.
\newblock Another subexponential-time quantum algorithm for the dihedral hidden
  subgroup problem.
\newblock 2011,  \href{http://xxx.lanl.gov/abs/1112.3333}{{\ttfamily
  arXiv:1112.3333}}.

\bibitem{lee:stateConversion}
T.~Lee, R.~Mittal, B.~W. Reichardt, R.~{\v Spalek}, and M.~Szegedy.
\newblock Quantum query complexity of the state conversion problem.
\newblock In {\em Proc. of 52nd IEEE FOCS}, pages 344--353, 2011,
  \href{http://xxx.lanl.gov/abs/1011.3020}{{\ttfamily arXiv:1011.3020}}.

\bibitem{magniez:walkSearch}
F.~Magniez, A.~Nayak, J.~Roland, and M.~Santha.
\newblock Search via quantum walk.
\newblock {\em SIAM Journal on Computing}, 40(1):142--164, 2011,
  \href{http://xxx.lanl.gov/abs/quant-ph/0608026}{{\ttfamily
  arXiv:quant-ph/0608026}}.

\bibitem{magniez:triangle}
F.~Magniez, M.~Santha, and M.~Szegedy.
\newblock Quantum algorithms for the triangle problem.
\newblock {\em SIAM Journal on Computing}, 37(2):413--424, 2007,
  \href{http://xxx.lanl.gov/abs/quant-ph/0310134}{{\ttfamily
  arXiv:quant-ph/0310134}}.

\bibitem{chuang:quantum}
M.~A. Nielsen and I.~L. Chuang.
\newblock {\em Quantum Computation and Quantum Information}.
\newblock Cambridge University Press, 2000.

\bibitem{reichardt:advTight}
B.~W. Reichardt.
\newblock Reflections for quantum query algorithms.
\newblock In {\em Proc. of 22nd ACM-SIAM SODA}, pages 560--569, 2011,
  \href{http://xxx.lanl.gov/abs/1005.1601}{{\ttfamily arXiv:1005.1601}}.

\bibitem{santha:walkBasedAlgorithms}
M.~Santha.
\newblock Quantum walk based search algorithms.
\newblock In {\em Proc. of 5th TAMC}, volume 4978 of {\em LNCS}, pages 31--46.
  Springer, 2008,  \href{http://xxx.lanl.gov/abs/0808.0059}{{\ttfamily
  arXiv:0808.0059}}.

\bibitem{schoning:ksat}
U.~Sch{\"o}ning.
\newblock A probabilistic algorithm for {$k$-SAT} and constraint satisfaction
  problems.
\newblock In {\em Proc. of 40th IEEE FOCS}, pages 410--414, 1999.

\bibitem{szegedy:walk}
M.~Szegedy.
\newblock Quantum speed-up of {Markov} chain based algorithms.
\newblock In {\em Proc. of 45th IEEE FOCS}, pages 32--41, 2004.

\end{thebibliography}

\end{document}